\documentclass[aps,prd,twocolumn,superscriptaddress,tightenlines,nofootinbib]{revtex4-1}
\usepackage[T1]{fontenc}
\usepackage{microtype}
\usepackage[textsize=scriptsize,backgroundcolor=red!70,linecolor=red]{todonotes}
\usepackage{booktabs}
\usepackage{dcolumn}
\usepackage{amsmath}
\usepackage{amsfonts}
\usepackage{amssymb}
\usepackage{graphicx}
\usepackage{ulem}
\usepackage{hyperref}
\usepackage[all]{hypcap}
\usepackage{paralist}
\usepackage{multirow}

\newcommand{\bs}[1]{\boldsymbol{#1}}
\usepackage{graphicx}
\usepackage{dcolumn}
\usepackage{bm}
\usepackage{epsf}
\usepackage{dcolumn}
\def\be{\begin{equation}}
\def\ee{\end{equation}}
\def\bea{\begin{eqnarray}}
\def\eea{\end{eqnarray}}

\def\beq{\begin{equation}}
\def\eeq{\end{equation}}
\def\d{\partial}
\def\beqa{\begin{eqnarray}}
\def\eeqa{\end{eqnarray}}

\begin{document}

\DeclareGraphicsExtensions{.eps,.ps}

\title{How to measure the reactor neutrino flux below the\\ inverse beta decay threshold with CE$\nu$NS}

\author{Jiajun Liao}
\email[Email Address: ]{liaojiajun@mail.sysu.edu.cn}
\affiliation{School of Physics, Sun Yat-sen University, Guangzhou, 510275, China}

\author{Hongkai Liu}
\email[Email Address: ]{liu.hongkai@campus.technion.ac.il}
\affiliation{Physics Department, Technion — Israel Institute of Technology, Haifa 3200003, Israel}
 
\author{Danny Marfatia}
\email[Email Address: ]{dmarf8@hawaii.edu}
\affiliation{Department of Physics and Astronomy, University of Hawaii at Manoa, Honolulu, HI 96822, USA}

\begin{abstract}
Most antineutrinos produced in a nuclear reactor have energies below the inverse beta decay threshold, and have not yet been detected.
We show that a coherent elastic neutrino-nucleus scattering experiment with an ultra-low energy threshold like NUCLEUS can measure the flux of reactor neutrinos below 1.8~MeV. 
Using a regularized unfolding procedure, we find that a meaningful upper bound can be placed on the low energy flux,
but the existence of the neutron capture component cannot be established. 
\end{abstract}
\maketitle

{\bf Introduction.} 
Nuclear reactors, as a steady and intense electron antineutrino source, have played a crucial role in the study of neutrino physics. In fact, neutrinos were first detected from a nuclear reactor via the inverse beta decay~(IBD) process $\bar\nu_e + p \to e^+ +  n$~\cite{Reines:1953pu,Reines:1959nc}. IBD has been commonly used to detect reactor antineutrinos in scintillators because 
the correlated signature of prompt positron emission followed by the delayed neutron capture significantly reduces backgrounds. However, since IBD has a 1.8~MeV threshold, the reactor antineutrino spectrum below this energy has not been measured.

There are two ways to calculate the reactor antineutrino spectrum: the beta spectrum conversion method and the {\it ab initio} summation method~\cite{Hayes:2016qnu}.
The conversion method uses the experimentally measured electron spectrum from a reactor core. Since no information on the fission yields and branching ratios is needed in the conversion method, it has better precision than the summation method. A limitation is that the measured electron data only allow an estimate of the reactor antineutrino spectrum between 2~MeV and 8~MeV~\cite{Mueller:2011nm,Huber:2011wv, Kopeikin:2021ugh}.

The summation method predicts the antineutrino spectrum by summing over all contributions of fission products from the nuclear data libraries, and can be used to calculate the antineutrino spectrum in a wide energy range. However, the  method is plagued by the {\it Pandemonium effect}~\cite{Hardy:1977suw} which arises from the limited efficiency of detecting gamma-rays from the de-excitation of high energy nuclear levels, and which leads to an overestimate of beta branching fractions of lower energy states in the nuclear databases. This in turn leads to an underestimate of the antineutrino flux below 2~MeV~\cite{Algora:2020mhh}. 

The reactor antineutrino flux above the IBD threshold has been well measured by large reactor neutrino experiments such as Daya Bay~\cite{DayaBay:2015lja}, RENO~\cite{RENO:2015ksa}, and Double Chooz~\cite{Crespo-Anadon:2014dea}.
If the reactor neutrino flux is measured below the IBD threshold, it will help refine calculations of the low energy antineutrino spectra using the summation method, which may help circumvent the Pandemonium effect.
Also, knowledge of the low-energy flux  may help understand if the assumed electron spectral shapes used to convert the measured aggregate electron spectrum (from the fission of each actinide) to the neutrino spectra are correct. This will reduce systematic uncertainties when converting the electron spectrum, which may shed light on the 5 MeV bump~\cite{DayaBay:2015lja,RENO:2015ksa,Crespo-Anadon:2014dea}.

Coherent elastic neutrino-nucleus scattering (CE$\nu$NS) occurs when the momentum transfer is smaller than the inverse radius of the nucleus. It was first observed by the COHERENT experiment in 2017 with a cesium iodide scintillation detector~\cite{COHERENT:2017ipa}. 
Recently, a measurement of CE$\nu$NS of reactor neutrinos above $\sim 6$~MeV was reported in Ref.~\cite{Colaresi:2022obx}. 
As reviewed in Ref.~\cite{Papoulias:2019xaw}, most studies have focused on CE$\nu$NS as a new tool for the study of neutrino properties and physics beyond the standard model using known source neutrino spectra. Instead, we propose to use the fact that CE$\nu$NS has no energy threshold to measure the reactor antineutrino flux below the IBD threshold. 
Although neutrino-electron elastic scattering also has no threshold, it can be neglected because its cross section is three orders of magnitude smaller than that of CE$\nu$NS~\cite{COHERENT:2017ipa}. 

The detection of low energy antineutrinos with CE$\nu$NS requires ultra-low threshold detectors. The NUCLEUS experiment~\cite{Strauss:2017cuu, NUCLEUS:2019igx}, which uses cryogenic detectors, may reach an unprecedented ${\cal{O}}(10)$~eV threshold and an ${\cal{O}}$(eV) energy resolution. The experiment has achieved a 20~eV threshold using a 0.5~g prototype made from 
Al$_2$O$_3$~\cite{Strauss:2017cuu}. In a phased multi-target approach,  a total 10~g mass of CaWO$_4$ and Al$_2$O$_3$ crystals, and eventually 1~kg of Ge, is foreseen. NUCLEUS-1kg, as a Ge detector, is expected to have a flat background with index below 100 counts/keV/kg/day and
an energy threshold of 5~eV~\cite{Strauss:2017cuu}. Although a flat background close to the threshold is overly optimistic, since Al$_2$O$_3$ has an order of magnitude smaller CE$\nu$NS signal than CaWO$_4$, effectively, it will measure the background for the signal in CaWO$_4$. As a result, we expect the background shape to be constrained by the time NUCLEUS-1kg starts taking data.
Since phonon energy will be measured, the suppression and uncertainty due to the nuclear quenching factor is eliminated~\cite{Liao:2021yog, Liao:2022hno}.
The detector will be placed on the surface at 72~m and 102~m from the two 4.25~GW reactor cores at the CHOOZ nuclear power plant. 
Cosmic ray induced events are the primary background. 

In this article, we show how to measure the reactor neutrino flux below the IBD threshold in a NUCLEUS-like CE$\nu$NS experiment.
The neutrino flux is obtained by unfolding the observed CE$\nu$NS spectrum.
Since simple unfolding produces highly oscillatory solutions,
it is necessary to impose some degree of smoothness on the unfolded distribution.  This injects bias into the solution. 
We perform regularized unfolding to minimize the amount of bias introduced.



\begin{figure}[t]
	\centering
	\includegraphics[width=0.4\textwidth]{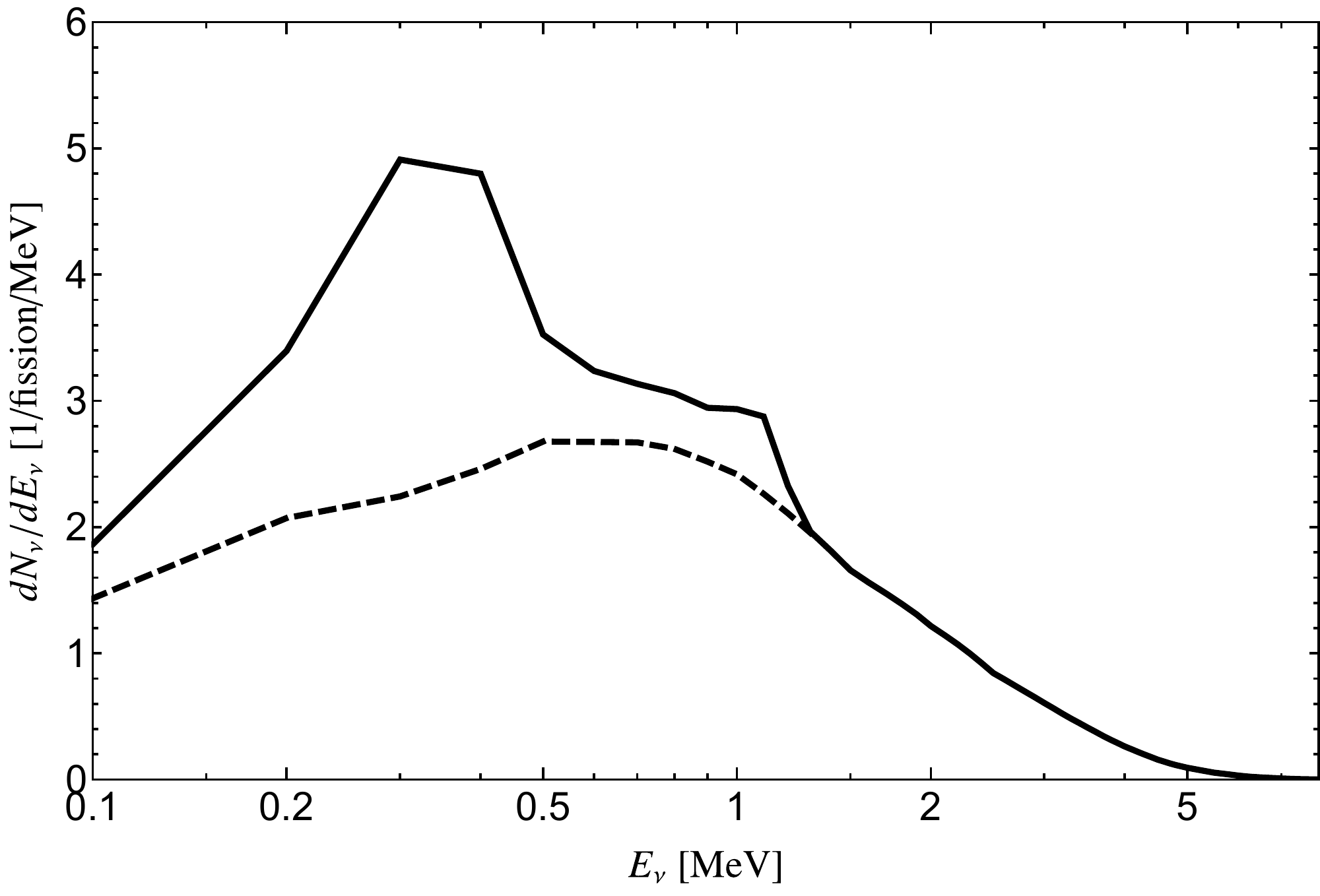}
	\caption{A typical prediction of the reactor antineutrino spectrum. The solid~(dashed) line corresponds to the case with~(without) the neutron capture component. }
	\label{fig:flux}
\end{figure}   

{\bf Reactor neutrino flux.} 
According to the summation method, roughly 84\% of the total neutrino flux from a commercial reactor arises from the beta decay of fission products of the principle fissile isotopes: $^{235}$U, $^{238}$U, $^{239}$Pu and $^{241}$Pu. 
The remaining 16\% comes from the capture of 0.61 neutrons/fission on $^{238}$U:  $^{238}$U + n $\to$ $^{239}$U $\to$ $^{239}$Np $\to$ $^{239}$Pu. As a result of the two $\beta$ decays, 1.22 
$\bar{\nu}_e$ are emitted per fission.
A prediction of the reactor neutrino flux per fission with~(without) the neutron capture component is shown by the solid~(dashed) line in Fig.~\ref{fig:flux}. Typical relative rates of $^{235}$U, $^{238}$U, $^{239}$Pu $^{241}$Pu (per fission) and the neutron capture process are given by $0.55: 0.07: 0.32: 0.06: 0.60$~\cite{TEXONO:2006xds}.
For the neutrino spectrum above 2~MeV, we adopt the predictions of the conversion method.
We use the results of Ref.~\cite{Kopeikin:2021ugh}
for $^{235}$U and $^{238}$U, which resolves the reactor antineutrino anomaly~\cite{Mention:2011rk} and is in a good agreement with the Daya Bay and RENO fuel evolution data~\cite{DayaBay:2019yxq,RENO:2018pwo}, and with STEREO data~\cite{STEREO:2022nzk}. For $^{239}$Pu and $^{241}$Pu we use the results of Ref.~\cite{Huber:2011wv}.
The fission components with energy below 2~MeV are taken from Ref.~\cite{Vogel:1989iv}, which is based on the summation method. 
The neutrino flux produced by neutron capture is derived from the standard beta spectra of $^{239}$U and $^{239}$Np~\cite{Kopeikin:1997ve,TEXONO:2006xds}, and has energies below about 1.3~MeV. 
As Fig.~\ref{fig:flux} suggests, roughly 70\% of the neutrinos have energies below the IBD threshold, and have not been experimentally detected. 
%

{\bf CE$\nu$NS spectrum.} 
The differential CE$\nu$NS event rate as a function of the nuclear recoil energy $E_R$ is
\beq
\frac{dN}{d E_R} = N_T \int \frac{d\Phi}{dE_\nu}\frac{d\sigma}{dE_R} d E_\nu\,,
\label{eq:dRdER}
\eeq
where $E_{\nu}$ is the antineutrino energy and $N_T$ is the number of nuclei in the detector. 
The reactor antineutrino flux is given by
\beq
\frac{d\Phi}{dE_\nu} = \frac{P}{\tilde{\epsilon} }\frac{1}{4\pi d_{\text{eff}}^2} \frac{d N_\nu}{dE_\nu}\,,
\eeq
where $P=4.25~\text{GW} \simeq 2.65\times 10^{22}~\text{MeV/second}$ is the reactor thermal power and $\tilde{\epsilon}=205.3$~MeV is the average energy released per fission. The effective distance $d_\text{eff}\equiv (1/d_1^2+ 1/d_2^2)^{-1/2}$ where $d_1=72$~m and $d_2=102$~m are the distances between the detector and reactor cores. 
The reactor neutrino spectrum per fission is $\frac{dN_\nu}{dE_\nu}$.
The standard model CE$\nu$NS cross section is given by~\cite{Freedman:1973yd}
\begin{equation}
\label{eq:crossSM}
\frac{d\sigma}{dE_R}=\frac{G_F^2M_N}{4\pi}q_W^2\left(1-\frac{M_NE_R}{2E_{\nu}^2}\right)F^2(\mathfrak{q})\,,
\end{equation}
where $M_N$ is the nuclear mass, $G_F$ is the Fermi coupling constant,  $q_W = N_n - (1 -4 \sin^2\theta_W)Z$ is the weak nuclear charge in terms of the weak mixing angle $\theta_W$, and $F(\mathfrak{q})$ is the nuclear form factor as a function of the momentum transfer $\mathfrak{q}$~\cite{Klein:1999gv}.
Due to the low momentum transfer in 
CE$\nu$NS with reactor antineutrinos, the predicted signal is not sensitive to the specific choice of the commonly used form factors and its uncertainties~\cite{AristizabalSierra:2019zmy}.

%
%
\begin{figure}[t]
\centering
\includegraphics[width=0.4\textwidth]{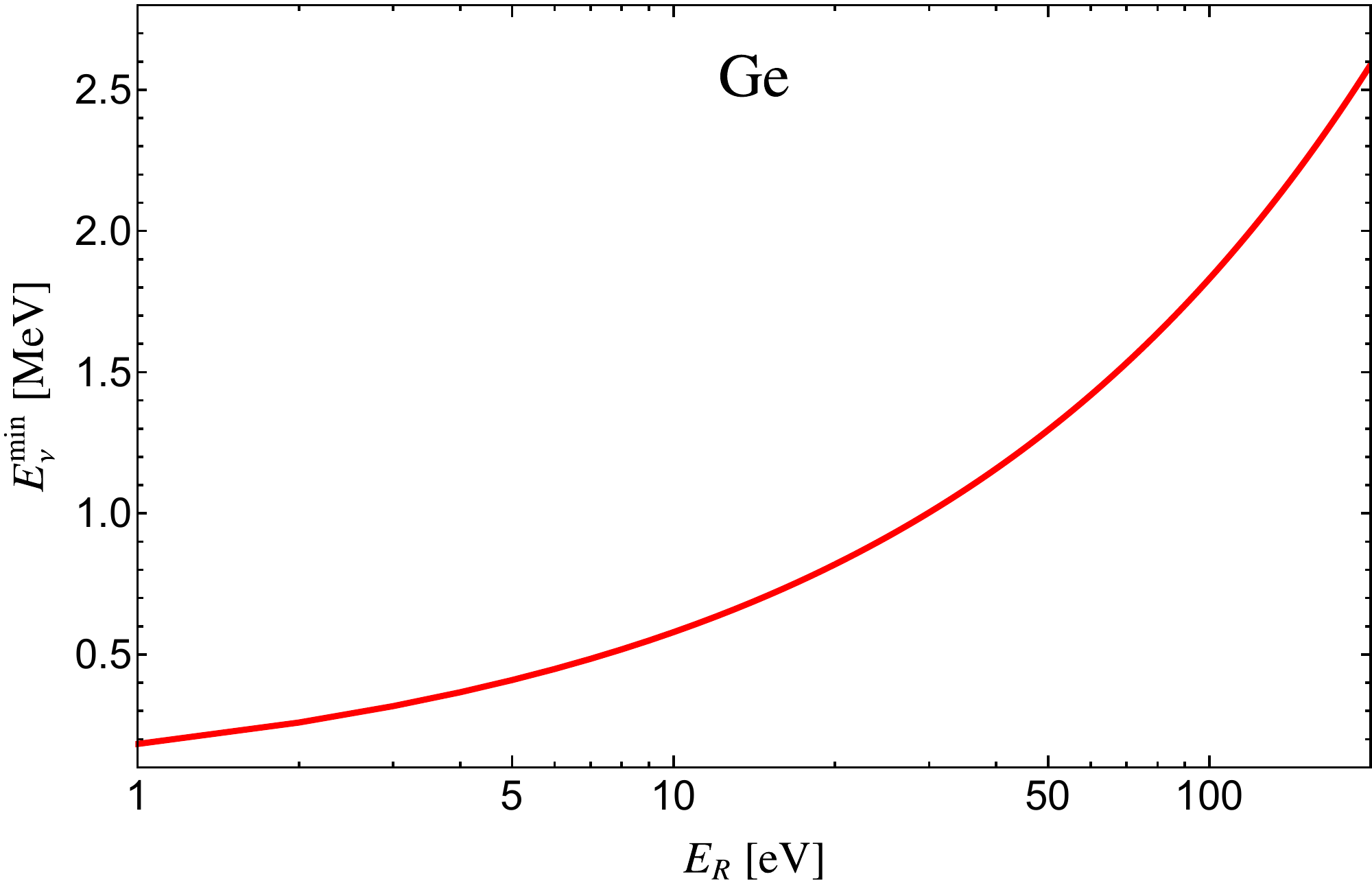}
	\caption{The minimum neutrino energy as a function of nuclear recoil energy $E_R$ in germanium.}
	\label{fig:thr}
\end{figure}

\begin{figure}[t]
\centering
\includegraphics[width=0.4\textwidth]{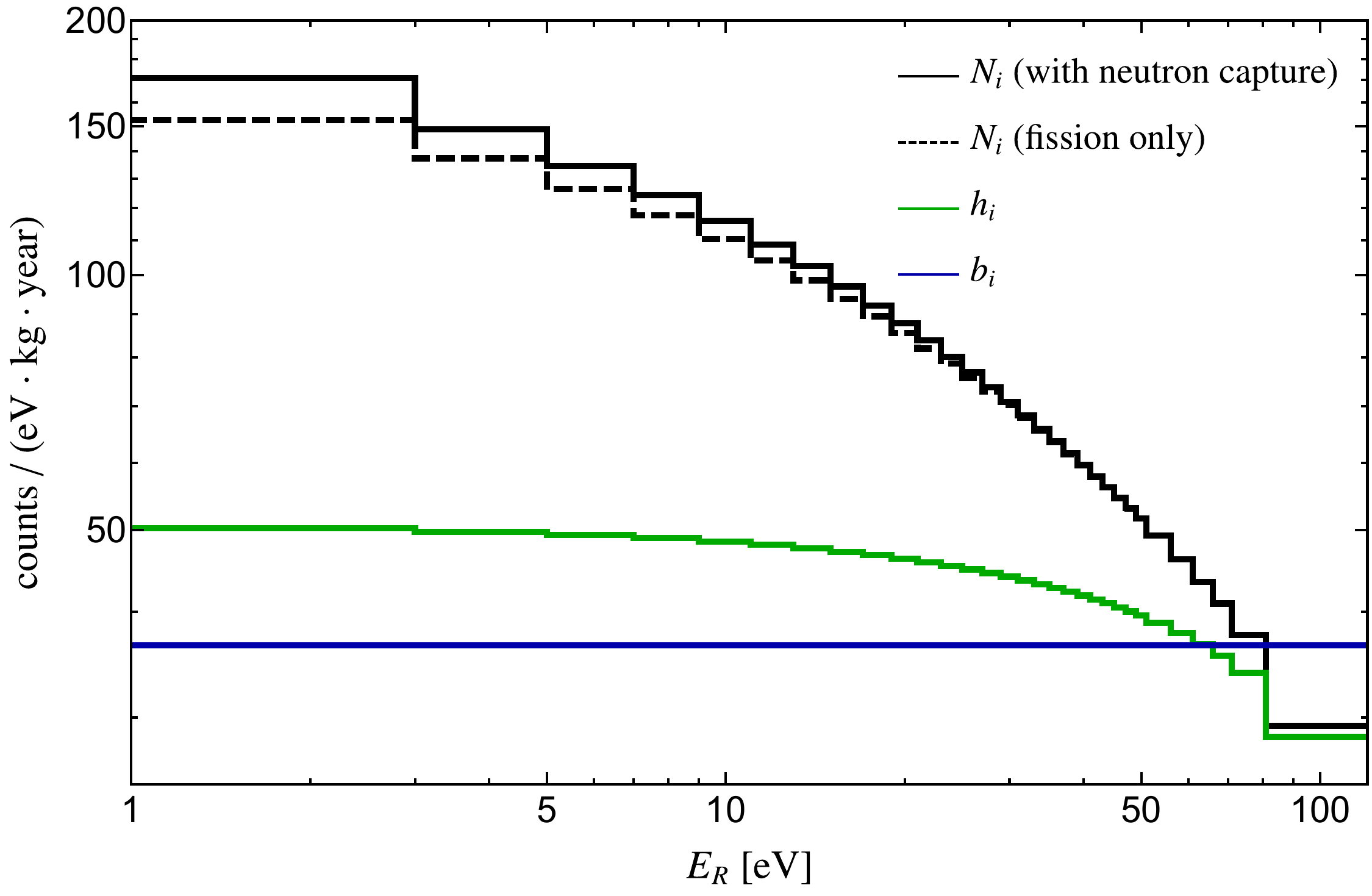}
	\caption{CE$\nu$NS spectra for the neutrino spectra in Fig.~\ref{fig:flux}. The green histogram corresponds to the contribution from the high-energy neutrino flux in the $j^{\text{th}}$ recoil energy bin, and the horizontal line corresponds to a flat background of 100 counts/keV/kg/day. 
 }
	\label{fig:CevNS}
\end{figure}

In exposure time $t$, the number of CE$\nu$NS events with recoil energy in the $j^{\rm th}$ bin $[E_R^j, E_R^{j+1}]$  is
\beq
N_j=t\int^{E_R^{j+1}}_{E_R^j}\frac{dN}{d E_R} dE_R\,.
\label{eq:counts}
\eeq
The minimum neutrino energy required to produce a nuclear recoil energy $E_R$ in Ge is shown in Fig.~\ref{fig:thr}.
With a threshold $E_{R, \rm{thr}}=5~(1)$~eV, only neutrinos with energy above 0.41~(0.18)~MeV can be detected.  Events with $E_R$ above 120~eV are determined by the neutrino flux above 2~MeV and are decoupled from the neutrino flux below 2~MeV. 
Since we assume that the flux is well known above 2~MeV, we need only consider nuclear recoil energies below 120~eV.
To study the neutrino flux below 
2~MeV, we use $m = 29~(31)$ bins between $E_{R, \rm{thr}} = 5~(1)$~eV and $E_R=120$~eV. 
The CE$\nu$NS spectra in a Ge detector corresponding to the antineutrino spectra in Fig.~\ref{fig:flux} are shown in
the upper panel of Fig.~\ref{fig:CevNS}.
The relatively large deviations in the bins with small recoil energy arise from the neutron capture contribution. 
 
 We consider the following three scenarios assuming a 1~eV energy resolution:
 {\it scenario 1}: $t = 1~\text{kg}\cdot \text{year},~\text{bkg} = 100{\ \rm{counts}}/(\text{keV}\cdot \text{kg}\cdot\text{day}),~E_{R,\text{thr}} = 5~$eV; {\it scenario 2}: $t = 3~\text{kg}\cdot \text{year},~\text{bkg} = 1{\ \rm{count}}/(\text{keV}\cdot \text{kg}\cdot\text{day}),~E_{R,\text{thr}} = 5~$eV; {\it scenario 3}: $t = 300~\text{kg}\cdot \text{year},~\text{bkg} = 1{\ \rm{count}}/(\text{keV}\cdot \text{kg}\cdot\text{day}),~E_{R,\text{thr}} = 1~$eV.
Our default configuration is scenario~1 and the other scenarios are for future upgrades with higher exposure, smaller background, and lower energy threshold. Scenario~3 may be unrealistic in the foreseeable future and is chosen to illustrate how difficult it is to identify the neutron capture component of the flux.\footnote{In principle, to establish the existence
of the neutron capture component, hypothesis testing
using forward folding from the predicted CE$\nu$NS spectrum can be carried out, but this requires a more careful understanding of the background than currently available.} In scenario 3, there are 31 $E_R$ bins as the threshold is lower. Note that the effect of a 1~eV energy resolution can be neglected since the bins are much larger.

{\bf Unfolding.}
To extract the neutrino flux at low energy from the CE$\nu$NS spectrum, we first split the integral over neutrino energy in the range $[0,2]$~MeV into $m$ bins. Then Eq.~(\ref{eq:counts}) becomes
%
$N_j = R_{ji}\nu_i + h_j$,
where $h_j$ is the contribution from the high-energy neutrino flux ($E_\nu >2$~MeV), and we take the low-energy neutrino flux per fission $\nu_i \equiv \frac{d N_\nu}{dE_\nu}|_{E_\nu^i=E^{\text{min}}_\nu(E_R^i)}$ to be  constant inside each neutrino energy bin. 
The square response matrix is
\beq
R_{ji} \equiv \frac{t N_T P}{\tilde{4\pi d_{\text{eff}}^2\epsilon} }  \int_{E_R^j}^{E_R^{j+1}} dE_R  \int_{E_\nu^i}^{E_\nu^{i+1}} dE_\nu \frac{d\sigma}{dE_R}\,,
\eeq and the contribution from the high-energy neutrino flux in the $j^{\text{th}}$ recoil energy bin is
\beq
h_j \equiv \frac{t N_T P}{4\pi d_{\text{eff}}^2 \tilde{\epsilon} }  \int_{E_R^j}^{E_R^{j+1}} dE_R  \int_{E_\nu > 2.0~\text{MeV}}^{\infty} \frac{d N_\nu}{dE_\nu}dE_\nu\frac{d\sigma}{dE_R}\,.
\eeq
We assume that the high-energy neutrino flux is known with a 3\% uncertainty. We have checked that this has a negligible effect on the determination of the low-energy neutrino flux.

\begin{figure}[t]
\centering
\includegraphics[width=0.4\textwidth]{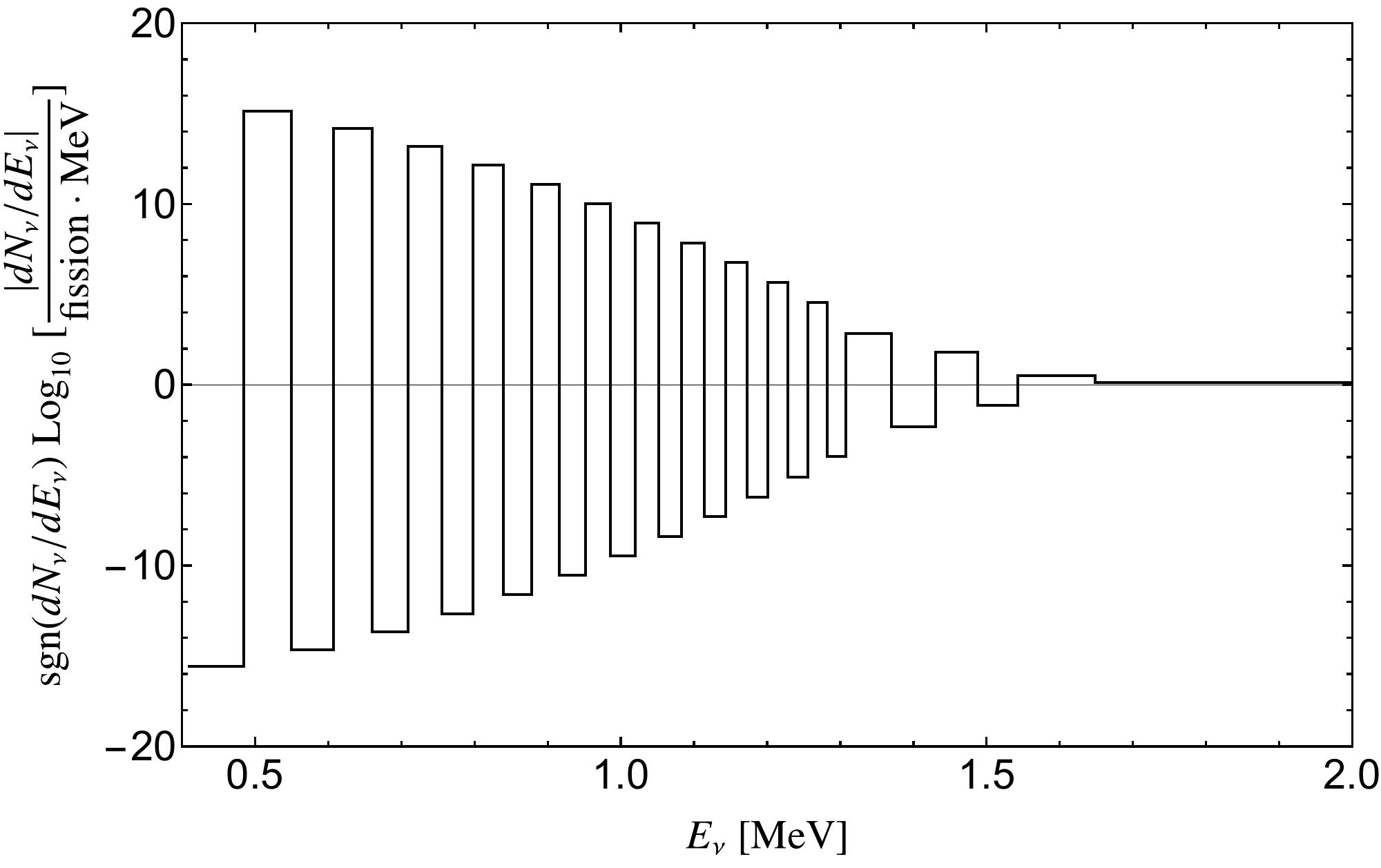}
	\caption{The neutrino flux obtained by simple unfolding. The true neutrino flux is the solid curve in Fig.~\ref{fig:flux} and coincides with the x-axis on the scale of this figure.}
	\label{fig:unfold}
\end{figure}

The observed number of events in the $j^{\text{th}}$ $E_R$ bin can be written as
\beq
\mu_j = R_{ji}\nu_i + h_j + b_j\,,
\label{eq:Nobs}
\eeq
where $b_j$ is the flat background. The spectrum of $h_j$ and $b_j$ are shown in Fig.~\ref{fig:CevNS}. 
We assume the actual number of events observed in each bin is $n_i$, where $n_i$ is an independent Poisson variable with expectation value $\mu_i = N_i + b_i$. 
Thus, the covariance matrix is 
$V_{ij} \equiv \text{cov}[n_i,n_j]= \delta_{ij}n_i$.
The low-energy $\bar{\nu}$ flux is easily solved by inverting Eq.~(\ref{eq:Nobs}):
\beq
\bs{\nu} = \bs{R}^{-1} (\bs{\mu}-\bs h-\bs b)\,.
\label{eq:unfold}
\eeq
We will refer to this as {\it simple} unfolding.
We take the estimator $\hat \mu_i$ to be $ n_i$ as it minimizes
\beq
\chi^2(\bs\nu) = \sum_{i=1}^m \frac{(\mu_i(\bs\nu)-n_i)^2}{n_i}\,,
\label{eq:chi2}
\eeq
where $\chi^2$ measures the significance with which the estimated CE$\nu$NS spectrum $\bs{\mu}$ deviates from the observed spectrum $\bs{n}$.
However, the unfolded flux $\bs \nu$ obtained
is highly oscillatory and takes negative values; see Fig.~\ref{fig:unfold}. The reason for these oscillations
is as follows~\cite{Cowan:1998ji}. The response matrix $\bs R$ smears fine structure in $\bs \nu$ and leaves some residual fine structure in $\bs \mu$. The effect of $\bs{R}^{-1}$ in Eq.~(\ref{eq:unfold}) is to restore this residual fine structure. Consequently, statistical fluctuations in the observed spectrum $\bs n$, which resemble residual fine structure, get amplified to
the oscillations in $\bs \nu$. Note that the amplitude
of the oscillations covers many orders of magnitude.
Because the unfolding procedure introduces large uncertainties, our unsophisticated modeling of the background is acceptable.

%

To ensure that the solution for the neutrino flux is smooth, we include a regularization function $S(\bs \nu)$ to define its smoothness. Instead of minimizing Eq.~(\ref{eq:chi2}), we minimize the regularized function~\cite{Cowan:1998ji},
\beq
\varphi(\bs\nu) = \chi^2(\bs\nu) + \beta S(\bs \nu)\,,
\eeq
 where $\beta$ is the regularization parameter and
 \beq
 S(\bs \nu) = \sum_{i=1}^{m-2} (-\nu_i+2\nu_{i+1}-\nu_{i+2})^2 = G_{ij}\nu_i\nu_j\,.
 \label{eq:regfun}
 \eeq
 This procedure is often called Tikhonov regularization~\cite{Tikhonov:1963an}. (We adopt the summation convention
 except for repeated indices denoting a diagonal element.)
The regularization function will take a large value if the neutrino flux solution has a large average curvature.  
By using the conditions,
 \beq
\frac{\d \varphi(\bs\nu)}{\d \nu_i} = D_{ij}\nu_j - K_j = 0\,,\quad i=1,2, \ldots m
\label{eq:cond}
\eeq
 we can solve for the unfolded neutrino flux $\hat{\bs\nu}$ analytically for a given $\bs n$:
\beq
\hat{\bs\nu}  =  \bs D^{-1} \bs K\,.
\label{eq:gen_unfold}
\eeq
Here,
\beq
{D_{ij}\over 2} = \frac{R_{ki}R_{kj}}{V_{kk}} +\beta G_{ij}\,,\ \ 
{K_i \over 2}= \frac{R_{ki}}{V_{kk}}(n_k-h_k-b_k)\,. 
\eeq
The estimated CE$\nu$NS spectrum $\bs{\hat\mu}$ can be computed by plugging the unfolded $\hat{\bs\nu}$ into Eq.~(\ref{eq:Nobs}): 
\beq
\hat{\bs\mu}(\beta,\bs{n}) = \bs R \,\hat{\bs\nu}(\beta,\bs{n}) +\bs h+\bs b\,.
\label{eq:sig}
\eeq
The neutrino flux obtained by minimizing the regularized function $\varphi$ will generally not give the minimum $\chi^2$. The deviation from the minimum $\chi^2$ value indicates that the model does not fit the data as well as it could. This 
is a consequence of neglecting part of the experimental information. 
The amount of information retained in a statistical model can be formulated using the regularization matrix~\cite{Cowan:1998ji},
\beq
M_{ij} \equiv \frac{\d\hat\mu_i}{\d n_j} = \frac{\d\hat\mu_i}{\d \hat{\nu}_k}\frac{\d\hat{\nu}_k}{\d n_j}=(\bs{RC})_{ij}\,,
\eeq
where
\beq
C_{ij}\equiv \frac{\d\hat\nu_i}{\d n_j} = 2(D^{-1})_{ik} {R_{jk} \over V_{jj}}\,.
\eeq
The reduced effective number of degrees of freedom, which quantifies the amount of experimental information ignored,  can be calculated by using the trace of the regularization matrix,
\beq
N_{\text{dof}} = m - \text{Tr}[\bs{M}]\,.
\label{eq:dof}
\eeq
Note that on setting $\beta = 0$, we retrieve the simple unfolding neutrino flux and the unbiased or least-$\chi^2$ estimator $\hat{\bs\mu} = \bs n$. 
To quantify the bias introduced by using the regularization function, we define the weighted sum of squares~\cite{Cowan:1998ji}
\beq
B = \sum_{i=1}^{m}\frac{\hat{b}^2_i}{W_{ii}}\,,
\label{eq:bias}
\eeq
where $\hat{b}_i \equiv \hat{\nu}_i(\beta, \boldsymbol \mu) - \nu_i \simeq \sum_{j=1}^m C_{ij} (\hat{\mu}_j - n_j)$ is the bias in $i^{\text{th}}$ bin, and the covariance matrix for the $\hat{b}_i$ is $\bs{W} = (\bs{CRC} - \bs C) \bs V (\bs{CRC} - \bs C)^T$. $B$ measures the deviation of the biases from zero.

Now we generate 3000 datasets by assuming a Poisson distribution in each bin,
\beq
n_i = \text{Poisson}(N_i + b_i),\quad i=1,2, \ldots m\,,
\eeq
where $N_i$ is either of the expected CE$\nu$NS spectra in Fig.~\ref{fig:CevNS}. We then repeat the following procedure for each dataset. (1)~Find the unfolded neutrino flux $\hat{\bs\nu}$ from Eq.~(\ref{eq:gen_unfold}). (2)~Obtain the expected CE$\nu$NS spectrum $\hat{\bs\mu}$ using Eq.~(\ref{eq:sig}). 
    (3)~Calculate $\chi^2$, $N_{\text{dof}}$, and $B$ based on Eq.~(\ref{eq:chi2}), Eq.~(\ref{eq:dof}), and Eq.~(\ref{eq:bias}), respectively. 

\begin{figure}
	\centering
	\includegraphics[width=0.4\textwidth]{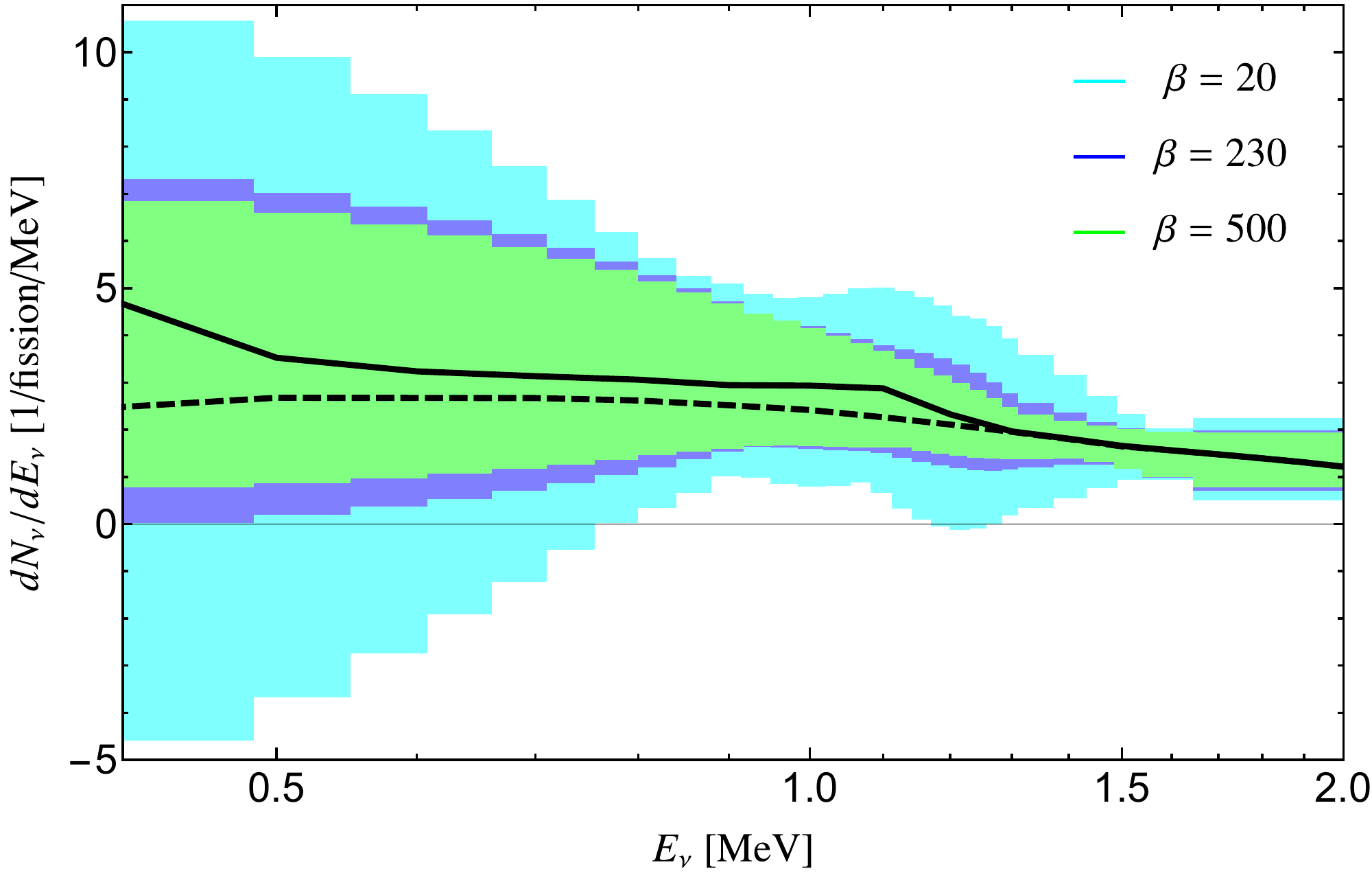}
	\caption{The 2$\sigma$ uncertainties in the determination of the low-energy reactor neutrino flux with different $\beta$ values for scenario~1. The black curves are the neutrino spectra in Fig.~\ref{fig:flux}. 
 The true neutrino flux is the solid curve which includes the  neutron capture component.}
	\label{fig:uncer}
\end{figure}

After carrying out this procedure we have 3000 values of $\chi^2$ and the corresponding $N_{\rm dof}$. To test how well the estimated CE$\nu$NS spectrum $\bs{\hat{\mu}}$ fits the observed CE$\nu$NS spectrum $\bs{n}$, we calculate the confidence level for each dataset for $N_{\text{dof}}$
with $\chi^2$ defined in Eq.~(\ref{eq:chi2}).
To make sure that $\bs{\hat{\mu}}$ is statistically compatible with $\bs{n}$, we only select those $\bs{\hat{\mu}}$ that fall within 2$\sigma$ and retain the corresponding neutrino flux $\bs{\hat{\nu}}$. Their envelope defines the 2$\sigma$ uncertainty in the neutrino flux.

\begin{figure}
	\centering
	\includegraphics[width=0.4\textwidth]{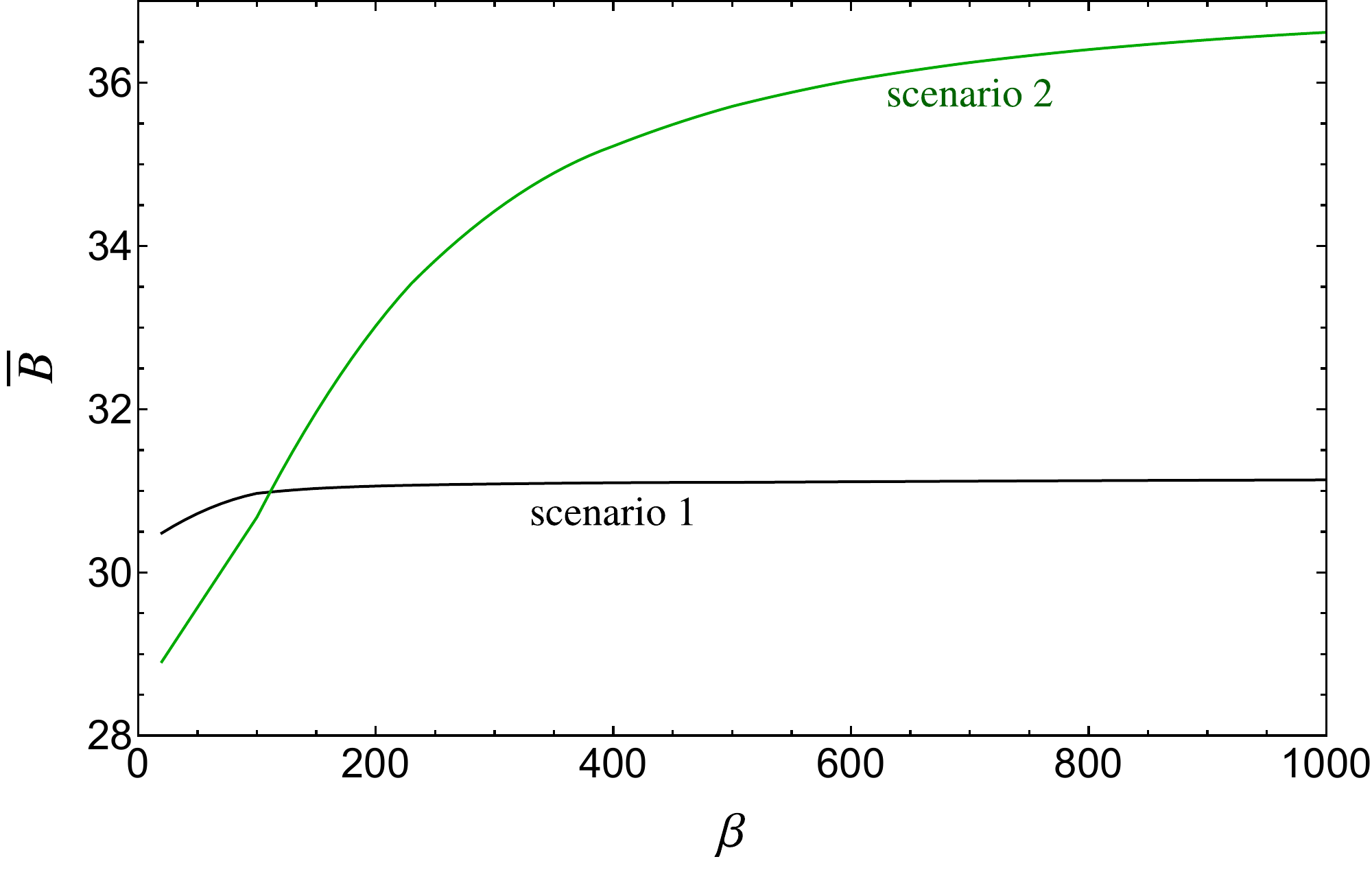}
	\caption{The average value of bias $\overline{B}$ as a function of $\beta$ for the first two scenarios.}
	\label{fig:bias}
\end{figure}

\begin{figure}[t]
	\centering
	\includegraphics[width=0.5\textwidth]{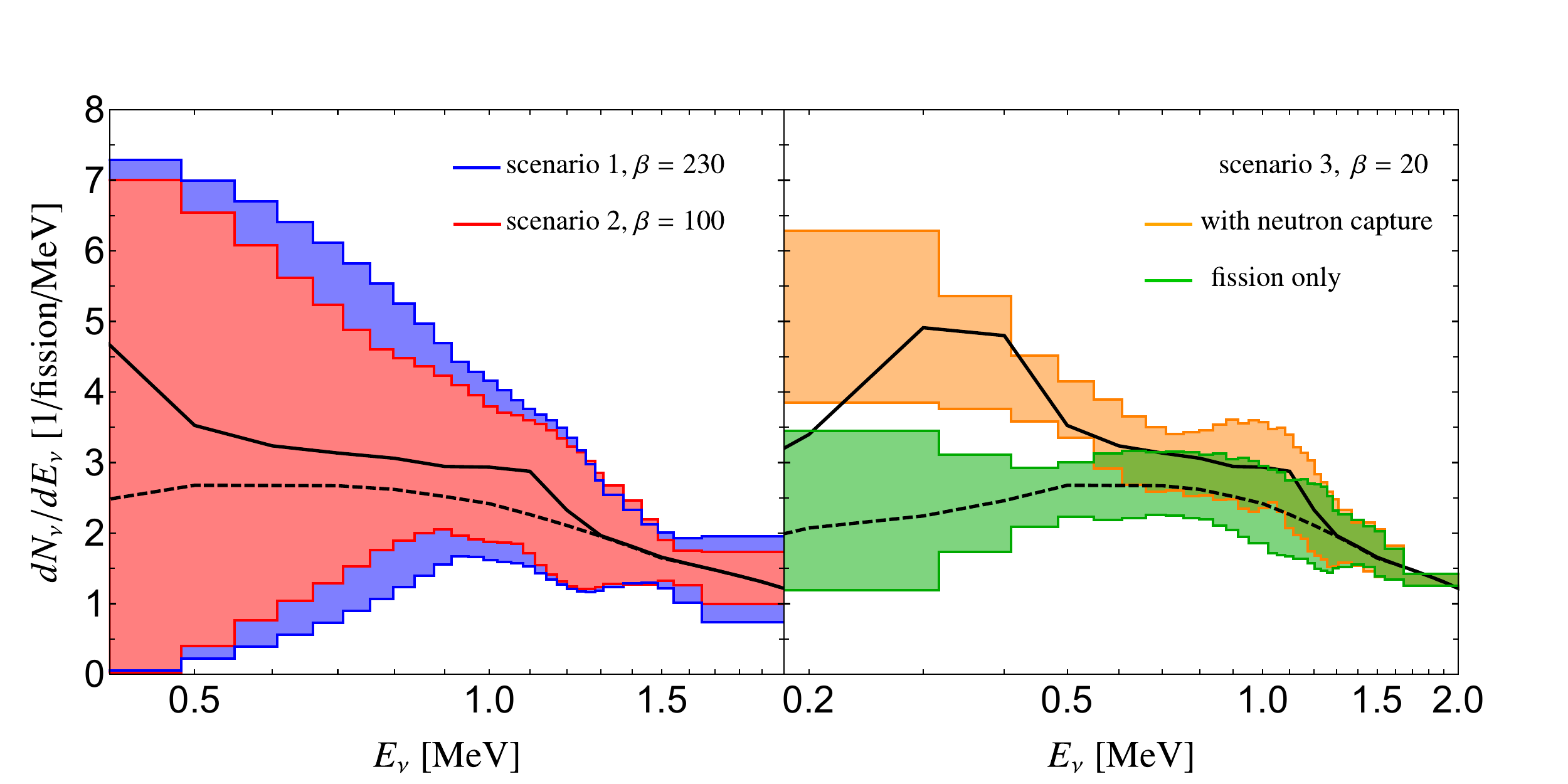}
	\caption{Left: The 2$\sigma$ uncertainties in the determination of the low-energy neutrino flux in scenarios 1 and 2. The true neutrino flux is the solid curve which includes the  neutron capture component. Right: The 2$\sigma$ uncertainties in the determination of the low-energy neutrino flux in scenario~3. The orange and green bands correspond to neutrino flux models with and without the neutron capture component, respectively.}
	\label{fig:future}
\end{figure}

We tried several values of $\beta$ from 20 to 5000. The $2\sigma$ uncertainties in the neutrino flux for different values of $\beta$ are shown in Fig.~\ref{fig:uncer} for scenario~1. As expected, larger $\beta$ suppresses the variance (at the expense of increased bias). 
Note from Fig.~\ref{fig:bias} that the average bias 
$\overline B$ plateaus 
at values that are not much larger than the number of bins $m$
(which is consistent with a strategy for selecting $\beta$ that lowers $\beta$ until $B \sim m$~\cite{Cowan:1998ji}). This means that a wide range of $\beta$ values works without introducing too much bias.
This is because a linear neutrino flux works very well for scenarios~1 and 2, which are not sensitive to $E_\nu$ below 0.41~MeV, and large values of $\beta$ force a linear neutrino flux. For all practical purposes, the $\beta \to \infty$ limit, at which the variance is minimized, is reached for $\beta \sim 1000$. 
 Figure~\ref{fig:uncer} shows that negative fluxes are permitted
 for $\beta=20$. As a criterion for selecting $\beta$, we choose the smallest value of $\beta$ that yields a positive definite flux at all energies. Accordingly, we fix $\beta = 230~(100)$ for scenario 1~(2). 
The $2\sigma$ uncertainty bands for scenarios~1 and 2 are shown in the left panel of Fig.~\ref{fig:future}. We see that the neutron capture component is buried in the uncertainties.

The reduced threshold of scenario~3 reveals the neutron capture 
bump more completely. 
To demonstrate the full capability of CE$\nu$NS, we show the $2\sigma$ neutrino flux band for this scenario in the right panel of Fig.~\ref{fig:future}. With such a large exposure, the variance is significantly reduced. 
The shape of the neutrino spectrum can be captured by using a smaller value of $\beta$ and correspondingly lower bias.
Note that our physical criterion that $\beta$ be selected to give a positive flux allows smaller values of $\beta$, in which case the uncertainty bands will have considerable overlap. 
In this sense, the result in the right panel of
Fig.~\ref{fig:future} is not robust, and is only illustrative.
A more restrictive criterion for the choice of $\beta$ needs to be devised. The question of the viability of the experiment envisioned in scenario~3 is more important.

{\bf Summary.} 
Most of the reactor neutrino flux has not been detected because it lies below the IBD threshold. Measuring the low energy neutrino flux can help improve theoretical models of the reactor neutrino spectrum. 
We assessed the potential of a NUCLEUS-like CE$\nu$NS experiment to measure the reactor neutrino flux below 2~MeV. 
A regularized unfolding procedure can be used to place an upper bound on the low energy flux with achievable experimental improvements.

{\it Acknowledgments.} 
J.L. is supported by the NNSF of China under Grant No. 12275368. H.L. is supported by the ISF, BSF and the Azrieli Foundation. D.M. is supported by the U.S. DoE under Grant No. de-sc0010504.

\twocolumngrid
\section*{References}
\def\bibsection{}
\bibliography{ref}

\begin{thebibliography}{30}%
\makeatletter
\providecommand \@ifxundefined [1]{%
 \@ifx{#1\undefined}
}%
\providecommand \@ifnum [1]{%
 \ifnum #1\expandafter \@firstoftwo
 \else \expandafter \@secondoftwo
 \fi
}%
\providecommand \@ifx [1]{%
 \ifx #1\expandafter \@firstoftwo
 \else \expandafter \@secondoftwo
 \fi
}%
\providecommand \natexlab [1]{#1}%
\providecommand \enquote  [1]{``#1''}%
\providecommand \bibnamefont  [1]{#1}%
\providecommand \bibfnamefont [1]{#1}%
\providecommand \citenamefont [1]{#1}%
\providecommand \href@noop [0]{\@secondoftwo}%
\providecommand \href [0]{\begingroup \@sanitize@url \@href}%
\providecommand \@href[1]{\@@startlink{#1}\@@href}%
\providecommand \@@href[1]{\endgroup#1\@@endlink}%
\providecommand \@sanitize@url [0]{\catcode `\\12\catcode `\$12\catcode
  `\&12\catcode `\#12\catcode `\^12\catcode `\_12\catcode `\%12\relax}%
\providecommand \@@startlink[1]{}%
\providecommand \@@endlink[0]{}%
\providecommand \url  [0]{\begingroup\@sanitize@url \@url }%
\providecommand \@url [1]{\endgroup\@href {#1}{\urlprefix }}%
\providecommand \urlprefix  [0]{URL }%
\providecommand \Eprint [0]{\href }%
\providecommand \doibase [0]{http://dx.doi.org/}%
\providecommand \selectlanguage [0]{\@gobble}%
\providecommand \bibinfo  [0]{\@secondoftwo}%
\providecommand \bibfield  [0]{\@secondoftwo}%
\providecommand \translation [1]{[#1]}%
\providecommand \BibitemOpen [0]{}%
\providecommand \bibitemStop [0]{}%
\providecommand \bibitemNoStop [0]{.\EOS\space}%
\providecommand \EOS [0]{\spacefactor3000\relax}%
\providecommand \BibitemShut  [1]{\csname bibitem#1\endcsname}%
\let\auto@bib@innerbib\@empty
\bibitem [{\citenamefont {Reines}\ and\ \citenamefont
  {Cowan}(1953)}]{Reines:1953pu}%
  \BibitemOpen
  \bibfield  {author} {\bibinfo {author} {\bibfnamefont {F.}~\bibnamefont
  {Reines}}\ and\ \bibinfo {author} {\bibfnamefont {C.~L.}\ \bibnamefont
  {Cowan}},\ }\href {\doibase 10.1103/PhysRev.92.830} {\bibfield  {journal}
  {\bibinfo  {journal} {Phys. Rev.}\ }\textbf {\bibinfo {volume} {92}},\
  \bibinfo {pages} {830} (\bibinfo {year} {1953})}\BibitemShut {NoStop}%
\bibitem [{\citenamefont {Reines}\ and\ \citenamefont
  {Cowan}(1959)}]{Reines:1959nc}%
  \BibitemOpen
  \bibfield  {author} {\bibinfo {author} {\bibfnamefont {F.}~\bibnamefont
  {Reines}}\ and\ \bibinfo {author} {\bibfnamefont {C.~L.}\ \bibnamefont
  {Cowan}},\ }\href {\doibase 10.1103/PhysRev.113.273} {\bibfield  {journal}
  {\bibinfo  {journal} {Phys. Rev.}\ }\textbf {\bibinfo {volume} {113}},\
  \bibinfo {pages} {273} (\bibinfo {year} {1959})}\BibitemShut {NoStop}%
\bibitem [{\citenamefont {Hayes}\ and\ \citenamefont
  {Vogel}(2016)}]{Hayes:2016qnu}%
  \BibitemOpen
  \bibfield  {author} {\bibinfo {author} {\bibfnamefont {A.~C.}\ \bibnamefont
  {Hayes}}\ and\ \bibinfo {author} {\bibfnamefont {P.}~\bibnamefont {Vogel}},\
  }\href {\doibase 10.1146/annurev-nucl-102115-044826} {\bibfield  {journal}
  {\bibinfo  {journal} {Ann. Rev. Nucl. Part. Sci.}\ }\textbf {\bibinfo
  {volume} {66}},\ \bibinfo {pages} {219} (\bibinfo {year} {2016})},\ \Eprint
  {http://arxiv.org/abs/1605.02047} {arXiv:1605.02047 [hep-ph]} \BibitemShut
  {NoStop}%
\bibitem [{\citenamefont {Mueller}\ \emph {et~al.}(2011)\citenamefont {Mueller}
  \emph {et~al.}}]{Mueller:2011nm}%
  \BibitemOpen
  \bibfield  {author} {\bibinfo {author} {\bibfnamefont {T.~A.}\ \bibnamefont
  {Mueller}} \emph {et~al.},\ }\href {\doibase 10.1103/PhysRevC.83.054615}
  {\bibfield  {journal} {\bibinfo  {journal} {Phys. Rev. C}\ }\textbf {\bibinfo
  {volume} {83}},\ \bibinfo {pages} {054615} (\bibinfo {year} {2011})},\
  \Eprint {http://arxiv.org/abs/1101.2663} {arXiv:1101.2663 [hep-ex]}
  \BibitemShut {NoStop}%
\bibitem [{\citenamefont {Huber}(2011)}]{Huber:2011wv}%
  \BibitemOpen
  \bibfield  {author} {\bibinfo {author} {\bibfnamefont {P.}~\bibnamefont
  {Huber}},\ }\href {\doibase 10.1103/PhysRevC.85.029901} {\bibfield  {journal}
  {\bibinfo  {journal} {Phys. Rev. C}\ }\textbf {\bibinfo {volume} {84}},\
  \bibinfo {pages} {024617} (\bibinfo {year} {2011})},\ \bibinfo {note}
  {[Erratum: Phys.Rev.C 85, 029901 (2012)]},\ \Eprint
  {http://arxiv.org/abs/1106.0687} {arXiv:1106.0687 [hep-ph]} \BibitemShut
  {NoStop}%
\bibitem [{\citenamefont {Kopeikin}\ \emph {et~al.}(2021)\citenamefont
  {Kopeikin}, \citenamefont {Skorokhvatov},\ and\ \citenamefont
  {Titov}}]{Kopeikin:2021ugh}%
  \BibitemOpen
  \bibfield  {author} {\bibinfo {author} {\bibfnamefont {V.}~\bibnamefont
  {Kopeikin}}, \bibinfo {author} {\bibfnamefont {M.}~\bibnamefont
  {Skorokhvatov}}, \ and\ \bibinfo {author} {\bibfnamefont {O.}~\bibnamefont
  {Titov}},\ }\href {\doibase 10.1103/PhysRevD.104.L071301} {\bibfield
  {journal} {\bibinfo  {journal} {Phys. Rev. D}\ }\textbf {\bibinfo {volume}
  {104}},\ \bibinfo {pages} {L071301} (\bibinfo {year} {2021})},\ \Eprint
  {http://arxiv.org/abs/2103.01684} {arXiv:2103.01684 [nucl-ex]} \BibitemShut
  {NoStop}%
\bibitem [{\citenamefont {Hardy}\ \emph {et~al.}(1977)\citenamefont {Hardy},
  \citenamefont {Carraz}, \citenamefont {Jonson},\ and\ \citenamefont
  {Hansen}}]{Hardy:1977suw}%
  \BibitemOpen
  \bibfield  {author} {\bibinfo {author} {\bibfnamefont {J.~C.}\ \bibnamefont
  {Hardy}}, \bibinfo {author} {\bibfnamefont {L.~C.}\ \bibnamefont {Carraz}},
  \bibinfo {author} {\bibfnamefont {B.}~\bibnamefont {Jonson}}, \ and\ \bibinfo
  {author} {\bibfnamefont {P.~G.}\ \bibnamefont {Hansen}},\ }\href {\doibase
  10.1016/0370-2693(77)90223-4} {\bibfield  {journal} {\bibinfo  {journal}
  {Phys. Lett. B}\ }\textbf {\bibinfo {volume} {71}},\ \bibinfo {pages} {307}
  (\bibinfo {year} {1977})}\BibitemShut {NoStop}%
\bibitem [{\citenamefont {Algora}\ \emph {et~al.}(2021)\citenamefont {Algora},
  \citenamefont {Tain}, \citenamefont {Rubio}, \citenamefont {Fallot},\ and\
  \citenamefont {Gelletly}}]{Algora:2020mhh}%
  \BibitemOpen
  \bibfield  {author} {\bibinfo {author} {\bibfnamefont {A.}~\bibnamefont
  {Algora}}, \bibinfo {author} {\bibfnamefont {J.~L.}\ \bibnamefont {Tain}},
  \bibinfo {author} {\bibfnamefont {B.}~\bibnamefont {Rubio}}, \bibinfo
  {author} {\bibfnamefont {M.}~\bibnamefont {Fallot}}, \ and\ \bibinfo {author}
  {\bibfnamefont {W.}~\bibnamefont {Gelletly}},\ }\href {\doibase
  10.1140/epja/s10050-020-00316-4} {\bibfield  {journal} {\bibinfo  {journal}
  {Eur. Phys. J. A}\ }\textbf {\bibinfo {volume} {57}},\ \bibinfo {pages} {85}
  (\bibinfo {year} {2021})},\ \Eprint {http://arxiv.org/abs/2007.07918}
  {arXiv:2007.07918 [nucl-ex]} \BibitemShut {NoStop}%
\bibitem [{\citenamefont {An}\ \emph {et~al.}(2016)\citenamefont {An} \emph
  {et~al.}}]{DayaBay:2015lja}%
  \BibitemOpen
  \bibfield  {author} {\bibinfo {author} {\bibfnamefont {F.~P.}\ \bibnamefont
  {An}} \emph {et~al.} (\bibinfo {collaboration} {Daya Bay}),\ }\href {\doibase
  10.1103/PhysRevLett.116.061801} {\bibfield  {journal} {\bibinfo  {journal}
  {Phys. Rev. Lett.}\ }\textbf {\bibinfo {volume} {116}},\ \bibinfo {pages}
  {061801} (\bibinfo {year} {2016})},\ \bibinfo {note} {[Erratum:
  Phys.Rev.Lett. 118, 099902 (2017)]},\ \Eprint
  {http://arxiv.org/abs/1508.04233} {arXiv:1508.04233 [hep-ex]} \BibitemShut
  {NoStop}%
\bibitem [{\citenamefont {Choi}\ \emph {et~al.}(2016)\citenamefont {Choi} \emph
  {et~al.}}]{RENO:2015ksa}%
  \BibitemOpen
  \bibfield  {author} {\bibinfo {author} {\bibfnamefont {J.~H.}\ \bibnamefont
  {Choi}} \emph {et~al.} (\bibinfo {collaboration} {RENO}),\ }\href {\doibase
  10.1103/PhysRevLett.116.211801} {\bibfield  {journal} {\bibinfo  {journal}
  {Phys. Rev. Lett.}\ }\textbf {\bibinfo {volume} {116}},\ \bibinfo {pages}
  {211801} (\bibinfo {year} {2016})},\ \Eprint
  {http://arxiv.org/abs/1511.05849} {arXiv:1511.05849 [hep-ex]} \BibitemShut
  {NoStop}%
\bibitem [{\citenamefont {Crespo-Anad\'on}(2015)}]{Crespo-Anadon:2014dea}%
  \BibitemOpen
  \bibfield  {author} {\bibinfo {author} {\bibfnamefont {J.~I.}\ \bibnamefont
  {Crespo-Anad\'on}} (\bibinfo {collaboration} {Double Chooz}),\ }\href
  {\doibase 10.1016/j.nuclphysbps.2015.06.025} {\bibfield  {journal} {\bibinfo
  {journal} {Nucl. Part. Phys. Proc.}\ }\textbf {\bibinfo {volume} {265-266}},\
  \bibinfo {pages} {99} (\bibinfo {year} {2015})},\ \Eprint
  {http://arxiv.org/abs/1412.3698} {arXiv:1412.3698 [hep-ex]} \BibitemShut
  {NoStop}%
\bibitem [{\citenamefont {Akimov}\ \emph {et~al.}(2017)\citenamefont {Akimov}
  \emph {et~al.}}]{COHERENT:2017ipa}%
  \BibitemOpen
  \bibfield  {author} {\bibinfo {author} {\bibfnamefont {D.}~\bibnamefont
  {Akimov}} \emph {et~al.} (\bibinfo {collaboration} {COHERENT}),\ }\href
  {\doibase 10.1126/science.aao0990} {\bibfield  {journal} {\bibinfo  {journal}
  {Science}\ }\textbf {\bibinfo {volume} {357}},\ \bibinfo {pages} {1123}
  (\bibinfo {year} {2017})},\ \Eprint {http://arxiv.org/abs/1708.01294}
  {arXiv:1708.01294 [nucl-ex]} \BibitemShut {NoStop}%
\bibitem [{\citenamefont {Colaresi}\ \emph {et~al.}(2022)\citenamefont
  {Colaresi}, \citenamefont {Collar}, \citenamefont {Hossbach}, \citenamefont
  {Lewis},\ and\ \citenamefont {Yocum}}]{Colaresi:2022obx}%
  \BibitemOpen
  \bibfield  {author} {\bibinfo {author} {\bibfnamefont {J.}~\bibnamefont
  {Colaresi}}, \bibinfo {author} {\bibfnamefont {J.~I.}\ \bibnamefont
  {Collar}}, \bibinfo {author} {\bibfnamefont {T.~W.}\ \bibnamefont
  {Hossbach}}, \bibinfo {author} {\bibfnamefont {C.~M.}\ \bibnamefont {Lewis}},
  \ and\ \bibinfo {author} {\bibfnamefont {K.~M.}\ \bibnamefont {Yocum}},\
  }\href {\doibase 10.1103/PhysRevLett.129.211802} {\bibfield  {journal}
  {\bibinfo  {journal} {Phys. Rev. Lett.}\ }\textbf {\bibinfo {volume} {129}},\
  \bibinfo {pages} {211802} (\bibinfo {year} {2022})},\ \Eprint
  {http://arxiv.org/abs/2202.09672} {arXiv:2202.09672 [hep-ex]} \BibitemShut
  {NoStop}%
\bibitem [{\citenamefont {Papoulias}\ \emph {et~al.}(2019)\citenamefont
  {Papoulias}, \citenamefont {Kosmas},\ and\ \citenamefont
  {Kuno}}]{Papoulias:2019xaw}%
  \BibitemOpen
  \bibfield  {author} {\bibinfo {author} {\bibfnamefont {D.~K.}\ \bibnamefont
  {Papoulias}}, \bibinfo {author} {\bibfnamefont {T.~S.}\ \bibnamefont
  {Kosmas}}, \ and\ \bibinfo {author} {\bibfnamefont {Y.}~\bibnamefont
  {Kuno}},\ }\href {\doibase 10.3389/fphy.2019.00191} {\bibfield  {journal}
  {\bibinfo  {journal} {Front. in Phys.}\ }\textbf {\bibinfo {volume} {7}},\
  \bibinfo {pages} {191} (\bibinfo {year} {2019})},\ \Eprint
  {http://arxiv.org/abs/1911.00916} {arXiv:1911.00916 [hep-ph]} \BibitemShut
  {NoStop}%
\bibitem [{\citenamefont {Strauss}\ \emph {et~al.}(2017)\citenamefont {Strauss}
  \emph {et~al.}}]{Strauss:2017cuu}%
  \BibitemOpen
  \bibfield  {author} {\bibinfo {author} {\bibfnamefont {R.}~\bibnamefont
  {Strauss}} \emph {et~al.},\ }\href {\doibase 10.1140/epjc/s10052-017-5068-2}
  {\bibfield  {journal} {\bibinfo  {journal} {Eur. Phys. J. C}\ }\textbf
  {\bibinfo {volume} {77}},\ \bibinfo {pages} {506} (\bibinfo {year} {2017})},\
  \Eprint {http://arxiv.org/abs/1704.04320} {arXiv:1704.04320
  [physics.ins-det]} \BibitemShut {NoStop}%
\bibitem [{\citenamefont {Angloher}\ \emph {et~al.}(2019)\citenamefont
  {Angloher} \emph {et~al.}}]{NUCLEUS:2019igx}%
  \BibitemOpen
  \bibfield  {author} {\bibinfo {author} {\bibfnamefont {G.}~\bibnamefont
  {Angloher}} \emph {et~al.} (\bibinfo {collaboration} {NUCLEUS}),\ }\href
  {\doibase 10.1140/epjc/s10052-019-7454-4} {\bibfield  {journal} {\bibinfo
  {journal} {Eur. Phys. J. C}\ }\textbf {\bibinfo {volume} {79}},\ \bibinfo
  {pages} {1018} (\bibinfo {year} {2019})},\ \Eprint
  {http://arxiv.org/abs/1905.10258} {arXiv:1905.10258 [physics.ins-det]}
  \BibitemShut {NoStop}%
\bibitem [{\citenamefont {Liao}\ \emph {et~al.}(2021)\citenamefont {Liao},
  \citenamefont {Liu},\ and\ \citenamefont {Marfatia}}]{Liao:2021yog}%
  \BibitemOpen
  \bibfield  {author} {\bibinfo {author} {\bibfnamefont {J.}~\bibnamefont
  {Liao}}, \bibinfo {author} {\bibfnamefont {H.}~\bibnamefont {Liu}}, \ and\
  \bibinfo {author} {\bibfnamefont {D.}~\bibnamefont {Marfatia}},\ }\href
  {\doibase 10.1103/PhysRevD.104.015005} {\bibfield  {journal} {\bibinfo
  {journal} {Phys. Rev. D}\ }\textbf {\bibinfo {volume} {104}},\ \bibinfo
  {pages} {015005} (\bibinfo {year} {2021})},\ \Eprint
  {http://arxiv.org/abs/2104.01811} {arXiv:2104.01811 [hep-ph]} \BibitemShut
  {NoStop}%
\bibitem [{\citenamefont {Liao}\ \emph {et~al.}(2022)\citenamefont {Liao},
  \citenamefont {Liu},\ and\ \citenamefont {Marfatia}}]{Liao:2022hno}%
  \BibitemOpen
  \bibfield  {author} {\bibinfo {author} {\bibfnamefont {J.}~\bibnamefont
  {Liao}}, \bibinfo {author} {\bibfnamefont {H.}~\bibnamefont {Liu}}, \ and\
  \bibinfo {author} {\bibfnamefont {D.}~\bibnamefont {Marfatia}},\ }\href
  {\doibase 10.1103/PhysRevD.106.L031702} {\bibfield  {journal} {\bibinfo
  {journal} {Phys. Rev. D}\ }\textbf {\bibinfo {volume} {106}},\ \bibinfo
  {pages} {L031702} (\bibinfo {year} {2022})},\ \Eprint
  {http://arxiv.org/abs/2202.10622} {arXiv:2202.10622 [hep-ph]} \BibitemShut
  {NoStop}%
\bibitem [{\citenamefont {Wong}\ \emph {et~al.}(2007)\citenamefont {Wong} \emph
  {et~al.}}]{TEXONO:2006xds}%
  \BibitemOpen
  \bibfield  {author} {\bibinfo {author} {\bibfnamefont {H.~T.}\ \bibnamefont
  {Wong}} \emph {et~al.} (\bibinfo {collaboration} {TEXONO}),\ }\href {\doibase
  10.1103/PhysRevD.75.012001} {\bibfield  {journal} {\bibinfo  {journal} {Phys.
  Rev. D}\ }\textbf {\bibinfo {volume} {75}},\ \bibinfo {pages} {012001}
  (\bibinfo {year} {2007})},\ \Eprint {http://arxiv.org/abs/hep-ex/0605006}
  {arXiv:hep-ex/0605006} \BibitemShut {NoStop}%
\bibitem [{\citenamefont {Mention}\ \emph {et~al.}(2011)\citenamefont
  {Mention}, \citenamefont {Fechner}, \citenamefont {Lasserre}, \citenamefont
  {Mueller}, \citenamefont {Lhuillier}, \citenamefont {Cribier},\ and\
  \citenamefont {Letourneau}}]{Mention:2011rk}%
  \BibitemOpen
  \bibfield  {author} {\bibinfo {author} {\bibfnamefont {G.}~\bibnamefont
  {Mention}}, \bibinfo {author} {\bibfnamefont {M.}~\bibnamefont {Fechner}},
  \bibinfo {author} {\bibfnamefont {T.}~\bibnamefont {Lasserre}}, \bibinfo
  {author} {\bibfnamefont {T.~A.}\ \bibnamefont {Mueller}}, \bibinfo {author}
  {\bibfnamefont {D.}~\bibnamefont {Lhuillier}}, \bibinfo {author}
  {\bibfnamefont {M.}~\bibnamefont {Cribier}}, \ and\ \bibinfo {author}
  {\bibfnamefont {A.}~\bibnamefont {Letourneau}},\ }\href {\doibase
  10.1103/PhysRevD.83.073006} {\bibfield  {journal} {\bibinfo  {journal} {Phys.
  Rev. D}\ }\textbf {\bibinfo {volume} {83}},\ \bibinfo {pages} {073006}
  (\bibinfo {year} {2011})},\ \Eprint {http://arxiv.org/abs/1101.2755}
  {arXiv:1101.2755 [hep-ex]} \BibitemShut {NoStop}%
\bibitem [{\citenamefont {Adey}\ \emph {et~al.}(2019)\citenamefont {Adey} \emph
  {et~al.}}]{DayaBay:2019yxq}%
  \BibitemOpen
  \bibfield  {author} {\bibinfo {author} {\bibfnamefont {D.}~\bibnamefont
  {Adey}} \emph {et~al.} (\bibinfo {collaboration} {Daya Bay}),\ }\href
  {\doibase 10.1103/PhysRevLett.123.111801} {\bibfield  {journal} {\bibinfo
  {journal} {Phys. Rev. Lett.}\ }\textbf {\bibinfo {volume} {123}},\ \bibinfo
  {pages} {111801} (\bibinfo {year} {2019})},\ \Eprint
  {http://arxiv.org/abs/1904.07812} {arXiv:1904.07812 [hep-ex]} \BibitemShut
  {NoStop}%
\bibitem [{\citenamefont {Bak}\ \emph {et~al.}(2019)\citenamefont {Bak} \emph
  {et~al.}}]{RENO:2018pwo}%
  \BibitemOpen
  \bibfield  {author} {\bibinfo {author} {\bibfnamefont {G.}~\bibnamefont
  {Bak}} \emph {et~al.} (\bibinfo {collaboration} {RENO}),\ }\href {\doibase
  10.1103/PhysRevLett.122.232501} {\bibfield  {journal} {\bibinfo  {journal}
  {Phys. Rev. Lett.}\ }\textbf {\bibinfo {volume} {122}},\ \bibinfo {pages}
  {232501} (\bibinfo {year} {2019})},\ \Eprint
  {http://arxiv.org/abs/1806.00574} {arXiv:1806.00574 [hep-ex]} \BibitemShut
  {NoStop}%
\bibitem [{\citenamefont {Almaz\'an}\ \emph {et~al.}(2023)\citenamefont
  {Almaz\'an} \emph {et~al.}}]{STEREO:2022nzk}%
  \BibitemOpen
  \bibfield  {author} {\bibinfo {author} {\bibfnamefont {H.}~\bibnamefont
  {Almaz\'an}} \emph {et~al.} (\bibinfo {collaboration} {STEREO}),\ }\href
  {\doibase 10.1038/s41586-022-05568-2} {\bibfield  {journal} {\bibinfo
  {journal} {Nature}\ }\textbf {\bibinfo {volume} {613}},\ \bibinfo {pages}
  {257} (\bibinfo {year} {2023})},\ \Eprint {http://arxiv.org/abs/2210.07664}
  {arXiv:2210.07664 [hep-ex]} \BibitemShut {NoStop}%
\bibitem [{\citenamefont {Vogel}\ and\ \citenamefont
  {Engel}(1989)}]{Vogel:1989iv}%
  \BibitemOpen
  \bibfield  {author} {\bibinfo {author} {\bibfnamefont {P.}~\bibnamefont
  {Vogel}}\ and\ \bibinfo {author} {\bibfnamefont {J.}~\bibnamefont {Engel}},\
  }\href {\doibase 10.1103/PhysRevD.39.3378} {\bibfield  {journal} {\bibinfo
  {journal} {Phys. Rev. D}\ }\textbf {\bibinfo {volume} {39}},\ \bibinfo
  {pages} {3378} (\bibinfo {year} {1989})}\BibitemShut {NoStop}%
\bibitem [{\citenamefont {Kopeikin}\ \emph {et~al.}(1997)\citenamefont
  {Kopeikin}, \citenamefont {Mikaelyan},\ and\ \citenamefont
  {Sinev}}]{Kopeikin:1997ve}%
  \BibitemOpen
  \bibfield  {author} {\bibinfo {author} {\bibfnamefont {V.~I.}\ \bibnamefont
  {Kopeikin}}, \bibinfo {author} {\bibfnamefont {L.~A.}\ \bibnamefont
  {Mikaelyan}}, \ and\ \bibinfo {author} {\bibfnamefont {V.~V.}\ \bibnamefont
  {Sinev}},\ }\href@noop {} {\bibfield  {journal} {\bibinfo  {journal} {Phys.
  Atom. Nucl.}\ }\textbf {\bibinfo {volume} {60}},\ \bibinfo {pages} {172}
  (\bibinfo {year} {1997})}\BibitemShut {NoStop}%
\bibitem [{\citenamefont {Freedman}(1974)}]{Freedman:1973yd}%
  \BibitemOpen
  \bibfield  {author} {\bibinfo {author} {\bibfnamefont {D.~Z.}\ \bibnamefont
  {Freedman}},\ }\href {\doibase 10.1103/PhysRevD.9.1389} {\bibfield  {journal}
  {\bibinfo  {journal} {Phys. Rev. D}\ }\textbf {\bibinfo {volume} {9}},\
  \bibinfo {pages} {1389} (\bibinfo {year} {1974})}\BibitemShut {NoStop}%
\bibitem [{\citenamefont {Klein}\ and\ \citenamefont
  {Nystrand}(2000)}]{Klein:1999gv}%
  \BibitemOpen
  \bibfield  {author} {\bibinfo {author} {\bibfnamefont {S.~R.}\ \bibnamefont
  {Klein}}\ and\ \bibinfo {author} {\bibfnamefont {J.}~\bibnamefont
  {Nystrand}},\ }\href {\doibase 10.1103/PhysRevLett.84.2330} {\bibfield
  {journal} {\bibinfo  {journal} {Phys. Rev. Lett.}\ }\textbf {\bibinfo
  {volume} {84}},\ \bibinfo {pages} {2330} (\bibinfo {year} {2000})},\ \Eprint
  {http://arxiv.org/abs/hep-ph/9909237} {arXiv:hep-ph/9909237} \BibitemShut
  {NoStop}%
\bibitem [{\citenamefont {Aristizabal~Sierra}\ \emph
  {et~al.}(2019)\citenamefont {Aristizabal~Sierra}, \citenamefont {Liao},\ and\
  \citenamefont {Marfatia}}]{AristizabalSierra:2019zmy}%
  \BibitemOpen
  \bibfield  {author} {\bibinfo {author} {\bibfnamefont {D.}~\bibnamefont
  {Aristizabal~Sierra}}, \bibinfo {author} {\bibfnamefont {J.}~\bibnamefont
  {Liao}}, \ and\ \bibinfo {author} {\bibfnamefont {D.}~\bibnamefont
  {Marfatia}},\ }\href {\doibase 10.1007/JHEP06(2019)141} {\bibfield  {journal}
  {\bibinfo  {journal} {JHEP}\ }\textbf {\bibinfo {volume} {06}},\ \bibinfo
  {pages} {141} (\bibinfo {year} {2019})},\ \Eprint
  {http://arxiv.org/abs/1902.07398} {arXiv:1902.07398 [hep-ph]} \BibitemShut
  {NoStop}%
\bibitem [{\citenamefont {Cowan}(1998)}]{Cowan:1998ji}%
  \BibitemOpen
  \bibfield  {author} {\bibinfo {author} {\bibfnamefont {G.}~\bibnamefont
  {Cowan}},\ }\href@noop {} {\emph {\bibinfo {title} {{Statistical data
  analysis}}}}\ (\bibinfo {year} {1998})\BibitemShut {NoStop}%
\bibitem [{\citenamefont {Tikhonov}(1963)}]{Tikhonov:1963an}%
  \BibitemOpen
  \bibfield  {author} {\bibinfo {author} {\bibfnamefont {A.~N.}\ \bibnamefont
  {Tikhonov}},\ }\href@noop {} {\bibfield  {journal} {\bibinfo  {journal} {Sov.
  Math.}\ }\textbf {\bibinfo {volume} {5}},\ \bibinfo {pages} {1035} (\bibinfo
  {year} {1963})}\BibitemShut {NoStop}%
\end{thebibliography}%

\end{document}